\documentclass[a4paper,11pt]{article}
\pdfoutput=1 

\usepackage{jcappub} 
\usepackage{aas_macros} 

\usepackage[T1]{fontenc} 
\usepackage{mathtools}

\usepackage{array}
\newcolumntype{P}[1]{>{\centering\arraybackslash}p{#1}}
\newcolumntype{M}[1]{>{\centering\arraybackslash}m{#1}}
\usepackage{graphicx}
\usepackage[dvipsnames]{xcolor}
\usepackage{longtable}
\usepackage{url}
\usepackage{verbatim}
\usepackage[utf8]{inputenc}
\usepackage{makecell}
\usepackage{comment}
\usepackage{cleveref}
\crefname{section}{§}{§§}
\usepackage[T1]{fontenc} 
\graphicspath{ {Figures/} }

\title{\boldmath Revealing ultra-high-energy cosmic ray acceleration with multi-messenger observations of the nearby GRB 980425/SN 1998bw}

\author[a,b,c]{Nestor Mirabal}

\affiliation[a]{Mail Code 661, Astroparticle Physics Laboratory,
NASA Goddard Space Flight Center, Greenbelt, MD 20771,
USA}
\affiliation[b]{University of Maryland, Baltimore County, MD 21250, USA}
\affiliation[c]{Center for Research and Exploration in Space Science and Technology, NASA Goddard Space Flight Center, Greenbelt, MD 20771}

\emailAdd{nestor.r.mirabalbarrios@nasa.gov}

\abstract{
The origin of ultra-high energy cosmic rays (UHECRs) is 
one of the most mystifying issues in astroparticle physics. It has been suggested that gamma-ray bursts (GRBs) are excellent acceleration sites for cosmic rays. 
The propagation of UHECRs from the GRB host galaxy to the Earth should generate delayed secondary photons and neutrinos. Here we present a dedicated search for delayed UHECR and neutrino emission centered around the position of nearby GRB 980425/SN 1998bw. Located at a distance of 36.9 Mpc, GRB 980425/SN 1998bw is well within the Greisen–Zatsepin–Kuzmin (GZK) distance horizon. 
We find no evidence for UHECR or neutrino clustering around the GRB 980425/SN 1998bw position between 2004 and 2020. Under ideal propagation conditions, we propose that it might be possible to detect  an excess from delayed UHECRs  around GRB 980425/SN 1998bw within the next 100 years if the intergalactic magnetic field (IGMF)
strength is  $B \leq 3 \times 10^{-13}$ G.}
\begin{document}
\maketitle
\flushbottom

\section{Introduction}
Gamma-ray bursts (GRBs) have long been considered excellent acceleration sites of UHECRs \citep{vietri,waxman2}. 
Secondary neutrinos and photons should also form as UHECRs produced by the GRB interact with the photon background along the line of sight \citep{1969PhLB...28..423B,1979ApJ...228..919S,waxman1}. Unfortunately, without further experimental evidence, it is not yet possible to exclude alternative UHECR sources \citep{10.3389/fspas.2019.00023}. Therefore, we need to start thinking about actual signals that might reveal the UHECR source in multidecadal time-scales. 

In our view, the minimum features that seem necessary for
potential UHECR signals in our lifetime are as follows:

\begin{itemize}
    \item Energetics: Enter your favorite model here \citep{hillas,10.3389/fspas.2019.00023}. There is an enormous amount of UHECR theoretical models that can satisfy the Hillas condition \citep{hillas}. However, for each UHECR model, one can find one supporting publication  and at least three publications stating that a different model is better. All things being equal, on paper most theoretical models appear to meet the stringent conditions required for UHECR energetics. 
    \item Distance/Distribution: This is an actual physical limitation that cannot be easily altered. For UHECR energies above 20 EeV, any successful UHECR source population must lie within 200 Mpc (or even closer) in order to satisfy the Greisen–Zatsepin–Kuzmin (GZK) distance horizon requirements \citep{greisen,zatsepin}. 
    \item Time Delay: This condition is somewhat tied to distance, but it is also dictated by the strength and distribution of magnetic fields \citep{dermer}. While there is no agreement on the actual UHECR accelerator, there is broad consensus that the arrival time of UHECRs will be delayed by tens to thousands of years  \citep{2012ApJ...748....9T}.
     A near real-time signature would require a relatively weak IGMF. 
\end{itemize}

There is one GRB source that appears to meet all the three conditions for a potential UHECR signal over a lifetime,  namely GRB 980425/SN 1998bw \citep{soffita,tinney,galama,kulkarni}. GRB980425/SN 1998bw was the first event to directly link  core-collapse supernovae (SNe) and long-duration GRBs \citep{galama}.  We know exactly when it triggered the UHECR acceleration \citep{soffita}. The energetics for UHECR acceleration in GRB 980425/SN 1998bw can be achieved either within the GRB itself \citep{waxman2} or through  the emerging SN 1998bw supernova  \citep{2006ApJ...651L...5M,wang,2011NatCo...2..175C}. Another excellent feature of GRB 980425/SN 1998bw is that a stellar explosion can readily provide heavy nuclei for UHECR acceleration. 
At a distance of 36.9 Mpc, GRB 980425/SN 1998bw continues to hold the record for the smallest redshift of all known GRBs ($z = 0.0085$) and it lies well within the GZK distance horizon \citep{greisen,zatsepin}. In terms of time delay, we are approaching nearly 25 years since the discovery of GRB 980425/SN 1998bw. 

The other key component needed for an actual signal detection is the availability of multi-messenger facilities. Fortunately, these already exist
as a confluence of multi-messenger observatories has been operating over the past few years.
UHECR data has been collected in the southern hemisphere by the Pierre Auger Observatory (PAO) since 2004 \citep{2004NIMPA.523...50A} and 
by the Telescope Array (TA) experiment in the northern hemisphere since 2008 \citep{ABUZAYYAD201287}. Neutrinos have been collected by the 
IceCube Neutrino Observatory partially since 2008 and with the full 86-string design starting in May 2011 \citep{icecube1}.  The Advanced LIGO and Advanced Virgo gravitational wave detectors have been collecting data since 2015 and 2017 respectively \citep{2020LRR....23....3A}. The  Fermi Gamma-Ray Space Telescope has been surveying the MeV-TeV sky since 2008 \citep{atwood}.
With all these facilities working together, we are truly living in the multi-messenger  era.

Here, we present a search for UHECRs and neutrinos centered around the position of GRB 980425/SN 1998bw. 
The paper is structured as follows: Section~\ref{uhecrs} describes the UHECR search, Section~\ref{neutrinos} describes the neutrino search and Section \ref{discussion} presents discussion and conclusions.

\section{The Pierre Auger Observatory and the UHECR sample}\label{uhecrs}

The Pierre Auger Observatory \citep{2004NIMPA.523...50A} is located
near Malarg{\"u}e in Argentina, at a longitude of $69^{\circ}$.4 and a latitude of -35$^{\circ}$.2. Its Surface Detector (SD) array consists of 1600 water-Cherenkov stations spread over  an area of 3000 km$^{2}$ to
observe extensive air showers generated by UHECRs. With a duty cycle of nearly 100\%, PAO can capture
muons, electrons and photons reaching ground level. 

\begin{figure}
	\includegraphics[width=\columnwidth]{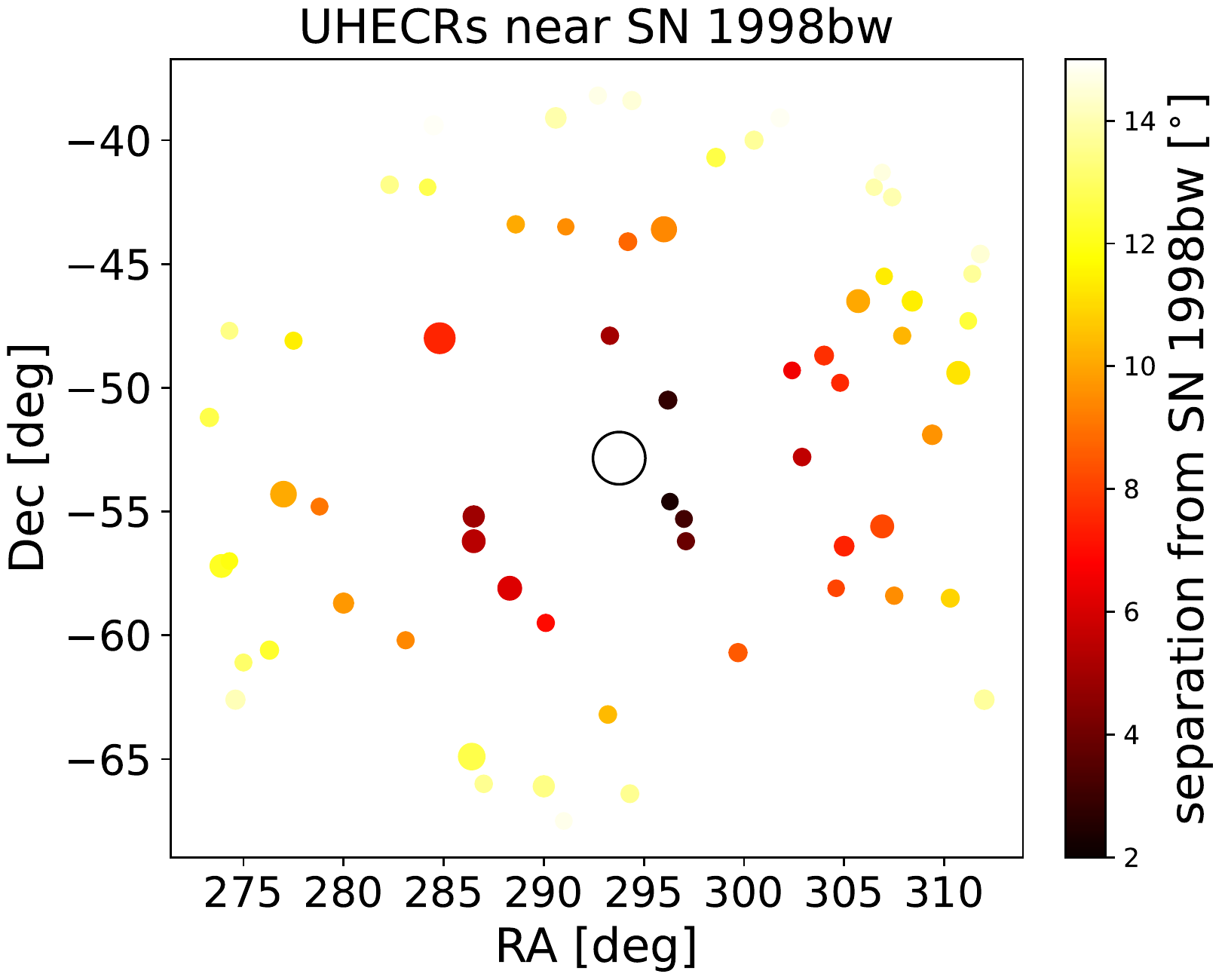}
    \caption{Positional scatter of PAO UHECRs near GRB 980425/SN 1998bw. The size of the filled marker is proportional to the cosmic-ray energy going from 32 EeV to 125 EeV. The open circle shown in the center of the image marks the position of GRB 980425/SN 1998bw.}
    \label{fig1}
\end{figure}

The data used in this paper corresponds to more than
2600 UHECR arrival directions above 32 EeV recorded with the SD array between
2004 January 1 to 2020 December 31 \citep{PAO}. For our analysis, we determined the 
angular separation between UHECR arrival directions and  the position of GRB 980425/SN 1998bw. 
For the GRB 980425/SN 1998bw position, we adopt reported measurements \citep{fynbo}. 
In Fig.~\ref{fig1} we show PAO UHECRs within a $5^{\circ}$ radius of  the GRB 980425/SN 1998bw position. 
Out of all
arrival directions, the nearest event to GRB 980425/SN 1998bw is a 32.9 EeV UHECR located at (RA, dec.) = (296$^{\circ}$.3, -54$^{\circ}$.6) 
 detected on 2010 January 27 with a deviation of 2$^{\circ}$.3  from
the position of GRB 980425/SN 1998bw.  

Following \cite{2014ApJ...794..172A,2019EPJWC.21001005B}, we searched for an UHECR excess using the Li and Ma method \citep{1983ApJ...272..317L} defining the  on-region using a 20-degree circular region around GRB 980425/SN 1998bw and the remainder of the the $3.4\pi$ steradians covered by PAO as the off-region  from which the background rate could be estimated. The 20-degree search region yields a $3.3\sigma$ significance centered at (RA, dec.) = (293$^{\circ}$.76,  -52$^{\circ}$.85). In the future, we plan to perform a more sensitive search looking for groups of directionally-aligned events or multiplets \citep{2009APh....32..269G,2012APh....35..354P}. To put the size of the 
analysis region in context, we remind the reader that 
the expected deflection angle for nuclei of charge $Z$                                      propagating a distance $D$ through an magnetic field $B$ and
a correlation length $l_{c}$ is given by $\theta_s \simeq 0.8^{\circ}$ $Z$ 
(D/10{\rm Mpc})$^{1/2}\, (l_{c}/1{\rm Mpc})^{1/2}\, (B/10^{-9}\, {\rm G})\,            
(E/10^{20}{\rm eV})^{-1} $
\citep{1996ApJ...462L..59M}. As a result, our UHECR excess search is most sensitive to protons and relatively light nuclei.

\section{IceCube Neutrinos}\label{neutrinos}
\begin{figure}
	\includegraphics[width=\columnwidth]{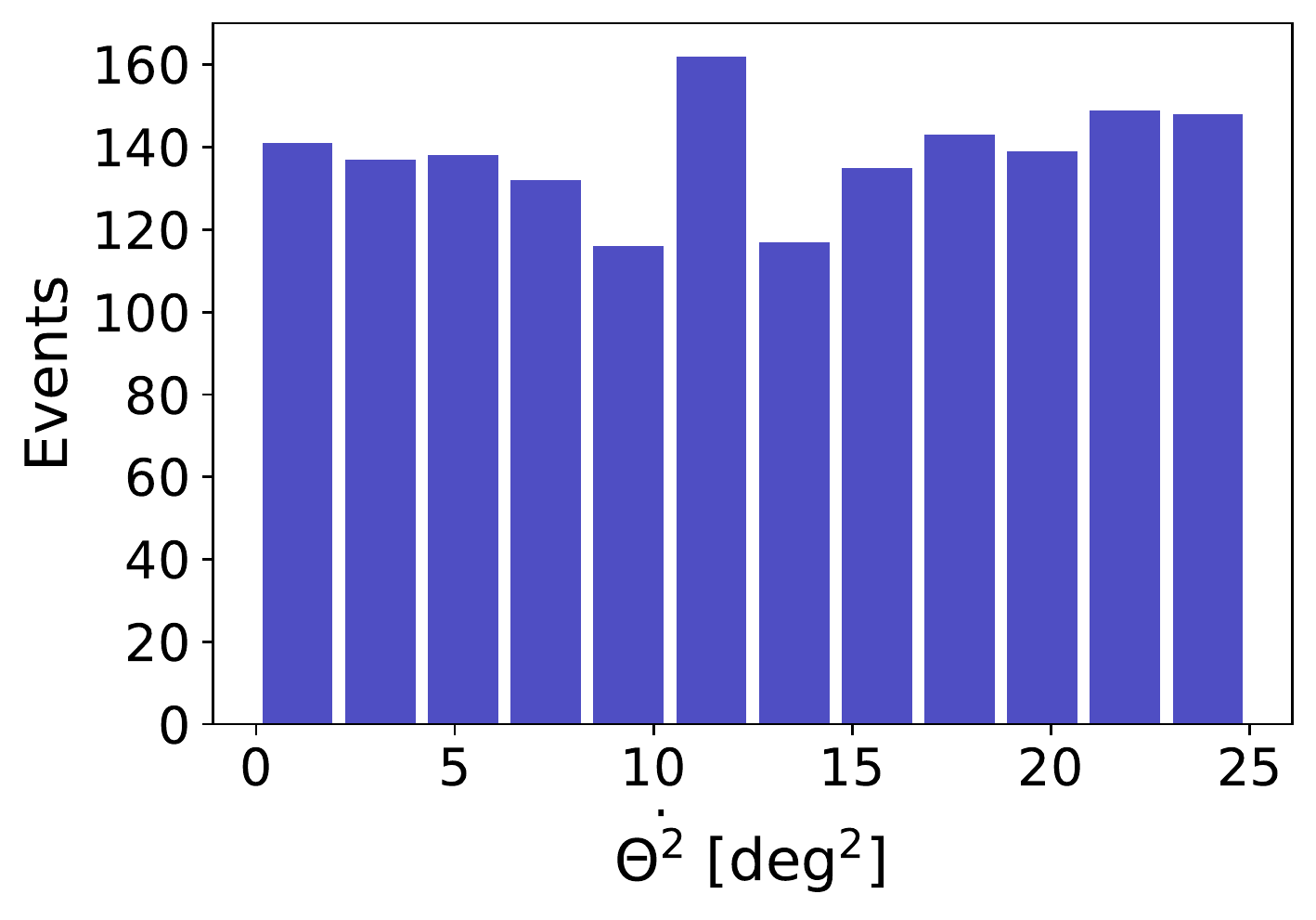}
    \caption{Binned distribution of the parameter $\Theta^2$, the squared angular distance between the individual neutrinos and the localization of GRB 980425/SN 1998bw. The background from athmospheric events clearly overwhelms any signal from GRB 980425/SN 1998bw.}
    \label{fig2}
\end{figure}

The IceCube Observatory is a 1 km$^{3}$ deep Cherenkov detector located in glacial ice near
the geographic South Pole \citep{achterberg2006first}. It is optimized to detect neutrinos above an energy threshold of $\sim$ 100 GeV. It consists of 5160 photomultiplier tubes (PMTs) along 86 strings. IceCube identifies neutrino interactions by tracking Cherenkov
light emitted by relativistic charged secondary particles
traveling through ice. We used 10 years of publicly released all-sky data recorded between 2008 April 6 and 2018 July 8 to search for neutrinos around the explosion site of GRB 980425/SN 1998bw \citep{icecube1}. The data includes events in the TeV-PeV range from partial configurations (40, 59, 79 strings) as well as the full 86-string configuration.
If pion production from UHECRs accelerated by GRBs takes place during the trajectory towards Earth there should be a measurable neutrino signal in the TeV-PeV range  \citep{2015NatCo...6.6783B,2022ApJ...939..116A}
In total, we collected 1134450 IceCube track-like events (see Table \ref{tab1}). We restricted our search to events within 5$^{\circ}$ of the GRB980425/SN 1998bw position, motivated by the recent IceCube detection of the nearby active galaxy NGC 1068 \citep{doi:10.1126/science.abg3395}. 
This narrows the total to 1657 events. Specifically, we analysed the distribution of the parameter $\theta^2$, the squared angular distance between individual neutrinos and the location of GRB980425/SN 1998bw. This is shown in Figure \ref{fig2}. 
Adopting a 1-degree signal region and a 5-degree background region yields a significance of 0.35$\sigma$ \citep{1983ApJ...272..317L}. 
Looking at the distribution it appears clearly dominated by background events from cosmic rays interacting with the atmosphere and producing athmospheric muons and neutrinos. Unfiltered, the athmospheric background  may be too strong to detect the underlying signal from nearby UHECR sources.

To further guide our search, we used a refined subset of high-energy neutrinos \citep{giommi2020dissecting}. The sample includes
70 well reconstructed and off the Galactic plane IceCube high-energy neutrinos  
recorded between 2009 August 13 and 2019 July 30  \citep{giommi2020dissecting}. The sample selected the best events from
IceCube’s diffuse astrophysical muon-neutrino search (DIF), high-energy starting tracks (HES) and alerts released through  IceCube’s realtime program (AHES, EHE).
Restricting the search to the Giommi et al. sample, we found that the closest match corresponds to IceCube-190124A located 22$^{\circ}$.9 from the GRB 980425/SN 1998bw explosion site. As discussed previously in Section 2, angular deflections restrict the neutrino search to UHECRs dominated by light nuclei.

\begin{table}
	\centering
	\caption{IceCube Data Set}
	\label{tab1}
	\begin{tabular}{lcr} 
		\hline
		  Sample & Livetime (days)  & Events (Number)\\
		\hline
            IC40 &	376.4	& 36900\\    
		IC59 & 352.6 & 107011\\
		IC79 & 316.0 & 93133\\
            IC86-I & 332.9 & 136244\\
		IC86-II &  332.0  & 112858\\
           IC86-III & 362.9 & 122541\\
            IC86-IV & 370.7 & 127045\\
            IC86-V & 365.4 & 129311\\
            IC86-VI & 357.3 & 123657\\
            IC86-VII & 410.6 & 145750\\ 
		\hline
	\end{tabular}
\end{table}

\section{Discussion and Conclusions}\label{discussion}

We have outlined three necessary conditions for finding a potential UHECR signal in multidecadal time-scales. We have further identified GRB 980425/SN 1998bw as the best available event to probe UHECR acceleration
in nearby GRBs
using multi-messenger facilities. 
Unfortunately, we find no obvious clustering of UHECRs or neutrinos within $5^{\circ}$ of the GRB 980425/SN 1998bw explosion site.

Assuming that a UHECR with energy $E_{p}$ and charge $Ze$ from GRB 980425/SN 1998bw 
travels a distance $d$ to reach us, we can use the time delay $\tau$ due to magnetic deflections to constrain the IGMF strength $B$ as a function of distance $d$ and the Larmor radius $r_{L}$. The expression for time delay of arrivals  gives \citep{dermer}

\begin{equation}
 \tau \simeq \frac{d^3}{24cr^2_L} \simeq
  \left(0.5 \frac{d^3_{3.5}Z^2B^{2}_{-12}}{E^{2}_{60}}\right)~{\rm days},
\end{equation}

\noindent
where $E_{60} = \frac{E_{p}}{60 EeV}$, $d_{3.5} = \frac{d} {3.5 Mpc}$,  $B_{-12} = \frac{B}{10^{-12} G}$ and electric charge $Z$.

At present, we have only studied the first 25 years of multi-messenger data since the GRB 980425/SN 1998bw explosion took place. The delay of UHECR arrivals could last tens to millions of years \citep{2012ApJ...748....9T}. If magnetic voids are abundant and the  IGMF strength is 
$B \leq 3 \times 10^{-13}$ G,
  we expect that an excess of UHECRs  might start to appear around the location of GRB 980425/SN 1998bw within the next few decades. 
  We note that there might be  additional time delays introduced by the UHECR transit within the GRB host galaxy and our own Galaxy, but we estimate such delays to be of order $\tau_{Galaxy} \simeq 4 B_{\mu G}^{2}~{\rm days}$ assuming $\sim$ kpc scales and $\mu$G magnetic field within the galaxies \cite{waxman1}.   Innovative UHECR and neutrino analyses of this part of the sky are highly encouraged. Delayed secondary photons from the electromagnetic cascades initiated by UHECRs/electron-positron pairs from GRB 980425/SN 1998bw may arrive much earlier than charged UHECRs. We will discuss delayed secondary gamma-ray emission from GRB 980425/SN 1998bw in a separate paper.

\acknowledgments
The material is based upon work supported by NASA under award number 80GSFC21M0002.
We thank the referee for useful comments that helped improve the paper. 
This research has made use of data obtained through the High Energy Astrophysics Science
Archive Research Center Online Service, provided by the NASA/Goddard Space Flight Center. This work made use of Astropy:\footnote{http://www.astropy.org} a community-developed core Python package and an ecosystem of tools and resources for astronomy.
All data used in this paper are publicly available.

\bibliographystyle{JHEP.bst}
\bibliography{refer.bib}

\providecommand{\href}[2]{#2}\begingroup\raggedright\begin{thebibliography}{10}

\bibitem{vietri}
M.~{Vietri}, \emph{{The Acceleration of Ultra--High-Energy Cosmic Rays in
  Gamma-Ray Bursts}}, \href{https://doi.org/10.1086/176448}{\emph{\apj}
  {\bfseries 453} (Nov., 1995) 883},
  [\href{https://arxiv.org/abs/astro-ph/9506081}{{\ttfamily
  astro-ph/9506081}}].

\bibitem{waxman2}
E.~{Waxman}, \emph{{Cosmological Origin for Cosmic Rays above 10 19 eV}},
  \href{https://doi.org/10.1086/309715}{\emph{\apjl} {\bfseries 452} (Oct.,
  1995) L1}, [\href{https://arxiv.org/abs/astro-ph/9508037}{{\ttfamily
  astro-ph/9508037}}].

\bibitem{1969PhLB...28..423B}
V.~S. {Beresinsky} and G.~T. {Zatsepin}, \emph{{Cosmic rays at ultra high
  energies (neutrino?)}},
  \href{https://doi.org/10.1016/0370-2693(69)90341-4}{\emph{Physics Letters B}
  {\bfseries 28} (Jan., 1969) 423--424}.

\bibitem{1979ApJ...228..919S}
F.~W. {Stecker}, \emph{{Diffuse fluxes of cosmic high-energy neutrinos.}},
  \href{https://doi.org/10.1086/156919}{\emph{\apj} {\bfseries 228} (Mar.,
  1979) 919--927}.

\bibitem{waxman1}
E.~{Waxman} and P.~{Coppi}, \emph{{Delayed GeV--TeV Photons from Gamma-Ray
  Bursts Producing High-Energy Cosmic Rays}},
  \href{https://doi.org/10.1086/310090}{\emph{\apjl} {\bfseries 464} (June,
  1996) L75}, [\href{https://arxiv.org/abs/astro-ph/9603144}{{\ttfamily
  astro-ph/9603144}}].

\bibitem{10.3389/fspas.2019.00023}
R.~Alves~Batista, J.~Biteau, M.~Bustamante, K.~Dolag, R.~Engel, K.~Fang et~al.,
  \emph{Open questions in cosmic-ray research at ultrahigh energies},
  \href{https://doi.org/10.3389/fspas.2019.00023}{\emph{Frontiers in Astronomy
  and Space Sciences} {\bfseries 6} (2019) }.

\bibitem{hillas}
A.~M. {Hillas}, \emph{{The Origin of Ultra-High-Energy Cosmic Rays}},
  \href{https://doi.org/10.1146/annurev.aa.22.090184.002233}{\emph{\araa}
  {\bfseries 22} (Jan., 1984) 425--444}.

\bibitem{greisen}
K.~Greisen, \emph{End to the cosmic-ray spectrum?},
  \href{https://doi.org/10.1103/PhysRevLett.16.748}{\emph{Phys. Rev. Lett.}
  {\bfseries 16} (Apr, 1966) 748--750}.

\bibitem{zatsepin}
G.~T. Zatsepin and V.~A. Kuz'min, \emph{Upper limit of the spectrum of cosmic
  rays}, {\emph{Soviet Journal of Experimental and Theoretical Physics Letters}
  {\bfseries 4} (1966) 78}.

\bibitem{dermer}
C.~D. {Dermer}, S.~{Razzaque}, J.~D. {Finke} and A.~{Atoyan},
  \emph{{Ultra-high-energy cosmic rays from black hole jets of radio
  galaxies}}, \href{https://doi.org/10.1088/1367-2630/11/6/065016}{\emph{New
  Journal of Physics} {\bfseries 11} (June, 2009) 065016},
  [\href{https://arxiv.org/abs/0811.1160}{{\ttfamily 0811.1160}}].

\bibitem{2012ApJ...748....9T}
H.~{Takami} and K.~{Murase}, \emph{{The Role of Structured Magnetic Fields on
  Constraining Properties of Transient Sources of Ultra-high-energy Cosmic
  Rays}}, \href{https://doi.org/10.1088/0004-637X/748/1/9}{\emph{\apj}
  {\bfseries 748} (Mar., 2012) 9},
  [\href{https://arxiv.org/abs/1110.3245}{{\ttfamily 1110.3245}}].

\bibitem{soffita}
P.~{Soffitta}, M.~{Feroci}, L.~{Piro}, J.~{in 't Zand}, J.~{Heise}, L.~{di
  Ciolo} et~al., \emph{{GRB 980425}}, {\emph{\iaucirc} {\bfseries 6884} (Apr.,
  1998) 1}.

\bibitem{tinney}
C.~{Tinney}, R.~{Stathakis}, R.~{Cannon}, T.~{Galama}, M.~{Wieringa}, D.~A.
  {Frail} et~al., \emph{{GRB 980425}}, {\emph{\iaucirc} {\bfseries 6896} (May,
  1998) 3}.

\bibitem{galama}
T.~J. {Galama}, P.~M. {Vreeswijk}, J.~{van Paradijs}, C.~{Kouveliotou},
  T.~{Augusteijn}, H.~{B{\"o}hnhardt} et~al., \emph{{An unusual supernova in
  the error box of the {\ensuremath{\gamma}}-ray burst of 25 April 1998}},
  \href{https://doi.org/10.1038/27150}{\emph{\nat} {\bfseries 395} (Oct., 1998)
  670--672}, [\href{https://arxiv.org/abs/astro-ph/9806175}{{\ttfamily
  astro-ph/9806175}}].

\bibitem{kulkarni}
S.~R. {Kulkarni}, D.~A. {Frail}, M.~H. {Wieringa}, R.~D. {Ekers}, E.~M.
  {Sadler}, R.~M. {Wark} et~al., \emph{{Radio emission from the unusual
  supernova 1998bw and its association with the {\ensuremath{\gamma}}-ray burst
  of 25 April 1998}}, \href{https://doi.org/10.1038/27139}{\emph{\nat}
  {\bfseries 395} (Oct., 1998) 663--669}.

\bibitem{2006ApJ...651L...5M}
K.~{Murase}, K.~{Ioka}, S.~{Nagataki} and T.~{Nakamura}, \emph{{High-Energy
  Neutrinos and Cosmic Rays from Low-Luminosity Gamma-Ray Bursts?}},
  \href{https://doi.org/10.1086/509323}{\emph{\apjl} {\bfseries 651} (Nov.,
  2006) L5--L8}, [\href{https://arxiv.org/abs/astro-ph/0607104}{{\ttfamily
  astro-ph/0607104}}].

\bibitem{wang}
X.-Y. Wang, S.~Razzaque, P.~M\'esz\'aros and Z.-G. Dai, \emph{High-energy
  cosmic rays and neutrinos from semirelativistic hypernovae},
  \href{https://doi.org/10.1103/PhysRevD.76.083009}{\emph{Phys. Rev. D}
  {\bfseries 76} (Oct, 2007) 083009}.

\bibitem{2011NatCo...2..175C}
S.~{Chakraborti}, A.~{Ray}, A.~M. {Soderberg}, A.~{Loeb} and P.~{Chandra},
  \emph{{Ultra-high-energy cosmic ray acceleration in engine-driven
  relativistic supernovae}},
  \href{https://doi.org/10.1038/ncomms1178}{\emph{Nature Communications}
  {\bfseries 2} (Feb., 2011) 175},
  [\href{https://arxiv.org/abs/1012.0850}{{\ttfamily 1012.0850}}].

\bibitem{2004NIMPA.523...50A}
J.~{Abraham}, M.~{Aglietta}, I.~C. {Aguirre}, M.~{Albrow}, D.~{Allard},
  I.~{Allekotte} et~al., \emph{{Properties and performance of the prototype
  instrument for the Pierre Auger Observatory}},
  \href{https://doi.org/10.1016/j.nima.2003.12.012}{\emph{Nuclear Instruments
  and Methods in Physics Research A} {\bfseries 523} (May, 2004) 50--95}.

\bibitem{ABUZAYYAD201287}
T.~Abu-Zayyad, R.~Aida, M.~Allen, R.~Anderson, R.~Azuma, E.~Barcikowski et~al.,
  \emph{The surface detector array of the telescope array experiment},
  \href{https://doi.org/https://doi.org/10.1016/j.nima.2012.05.079}{\emph{Nuclear
  Instruments and Methods in Physics Research Section A: Accelerators,
  Spectrometers, Detectors and Associated Equipment} {\bfseries 689} (2012)
  87--97}.

\bibitem{icecube1}
M.~G. {Aartsen}, K.~{Abraham}, M.~{Ackermann}, J.~{Adams}, J.~A. {Aguilar},
  M.~{Ahlers} et~al., \emph{{All-sky Search for Time-integrated Neutrino
  Emission from Astrophysical Sources with 7 yr of IceCube Data}},
  \href{https://doi.org/10.3847/1538-4357/835/2/151}{\emph{\apj} {\bfseries
  835} (Feb., 2017) 151}, [\href{https://arxiv.org/abs/1609.04981}{{\ttfamily
  1609.04981}}].

\bibitem{2020LRR....23....3A}
B.~P. {Abbott}, R.~{Abbott}, T.~D. {Abbott}, S.~{Abraham}, F.~{Acernese},
  K.~{Ackley} et~al., \emph{{Prospects for observing and localizing
  gravitational-wave transients with Advanced LIGO, Advanced Virgo and KAGRA}},
  \href{https://doi.org/10.1007/s41114-020-00026-9}{\emph{Living Reviews in
  Relativity} {\bfseries 23} (Sept., 2020) 3}.

\bibitem{atwood}
W.~B. {Atwood}, A.~A. {Abdo}, M.~{Ackermann}, W.~{Althouse}, B.~{Anderson},
  M.~{Axelsson} et~al., \emph{{The Large Area Telescope on the Fermi Gamma-Ray
  Space Telescope Mission}},
  \href{https://doi.org/10.1088/0004-637X/697/2/1071}{\emph{\apj} {\bfseries
  697} (June, 2009) 1071--1102},
  [\href{https://arxiv.org/abs/0902.1089}{{\ttfamily 0902.1089}}].

\bibitem{PAO}
{The Pierre Auger Collaboration}, P.~{Abreu}, M.~{Aglietta}, J.~M. {Albury},
  I.~{Allekotte}, K.~{Almeida Cheminant} et~al., \emph{{Arrival Directions of
  Cosmic Rays above 32 EeV from Phase One of the Pierre Auger Observatory}},
  {\emph{arXiv e-prints} (June, 2022) arXiv:2206.13492},
  [\href{https://arxiv.org/abs/2206.13492}{{\ttfamily 2206.13492}}].

\bibitem{fynbo}
J.~U. {Fynbo}, S.~{Holland}, M.~I. {Andersen}, B.~{Thomsen}, J.~{Hjorth},
  G.~{Bj{\"o}rnsson} et~al., \emph{{Hubble Space Telescope Space Telescope
  Imaging Spectrograph Imaging of the Host Galaxy of GRB 980425/SN 1998BW}},
  \href{https://doi.org/10.1086/312942}{\emph{\apjl} {\bfseries 542} (Oct.,
  2000) L89--L93}, [\href{https://arxiv.org/abs/astro-ph/0009014}{{\ttfamily
  astro-ph/0009014}}].

\bibitem{2014ApJ...794..172A}
A.~{Aab}, P.~{Abreu}, M.~{Aglietta}, E.~J. {Ahn}, I.~A. {Samarai}, I.~F.~M.
  {Albuquerque} et~al., \emph{{Searches for Large-scale Anisotropy in the
  Arrival Directions of Cosmic Rays Detected above Energy of {}10$^{19}$ eV at
  the Pierre Auger Observatory and the Telescope Array}},
  \href{https://doi.org/10.1088/0004-637X/794/2/172}{\emph{\apj} {\bfseries
  794} (Oct., 2014) 172}, [\href{https://arxiv.org/abs/1409.3128}{{\ttfamily
  1409.3128}}].

\bibitem{2019EPJWC.21001005B}
J.~{Biteau}, T.~{Bister}, L.~{Caccianiga}, O.~{Deligny}, A.~{di Matteo},
  T.~{Fujii} et~al., \emph{{Covering the celestial sphere at ultra-high
  energies: Full-sky cosmic-ray maps beyond the ankle and the flux
  suppression}},  in \emph{European Physical Journal Web of Conferences},
  vol.~210 of \emph{European Physical Journal Web of Conferences}, p.~01005,
  Oct., 2019, \href{https://arxiv.org/abs/1905.04188}{{\ttfamily 1905.04188}},
  \href{https://doi.org/10.1051/epjconf/201921001005}{DOI}.

\bibitem{1983ApJ...272..317L}
T.~P. {Li} and Y.~Q. {Ma}, \emph{{Analysis methods for results in gamma-ray
  astronomy.}}, \href{https://doi.org/10.1086/161295}{\emph{\apj} {\bfseries
  272} (Sept., 1983) 317--324}.

\bibitem{2009APh....32..269G}
G.~{Golup}, D.~{Harari}, S.~{Mollerach} and E.~{Roulet}, \emph{{Source position
  reconstruction and constraints on the galactic magnetic field from ultra-high
  energy cosmic rays}},
  \href{https://doi.org/10.1016/j.astropartphys.2009.09.003}{\emph{Astroparticle
  Physics} {\bfseries 32} (Dec., 2009) 269--277},
  [\href{https://arxiv.org/abs/0902.1742}{{\ttfamily 0902.1742}}].

\bibitem{2012APh....35..354P}
{Pierre Auger Collaboration}, P.~{Abreu}, M.~{Aglietta}, E.~J. {Ahn}, I.~F.~M.
  {Albuquerque}, D.~{Allard} et~al., \emph{{Search for signatures of
  magnetically-induced alignment in the arrival directions measured by the
  Pierre Auger Observatory}},
  \href{https://doi.org/10.1016/j.astropartphys.2011.10.004}{\emph{Astroparticle
  Physics} {\bfseries 35} (Jan., 2012) 354--361},
  [\href{https://arxiv.org/abs/1111.2472}{{\ttfamily 1111.2472}}].

\bibitem{1996ApJ...462L..59M}
J.~{Miralda-Escude} and E.~{Waxman}, \emph{{Signatures of the Origin of
  High-Energy Cosmic Rays in Cosmological Gamma-Ray Bursts}},
  \href{https://doi.org/10.1086/310042}{\emph{\apjl} {\bfseries 462} (May,
  1996) L59}, [\href{https://arxiv.org/abs/astro-ph/9601012}{{\ttfamily
  astro-ph/9601012}}].

\bibitem{achterberg2006first}
A.~Achterberg, M.~Ackermann, J.~Adams, J.~Ahrens, K.~Andeen, D.~Atlee et~al.,
  \emph{First year performance of the icecube neutrino telescope},
  {\emph{Astroparticle Physics} {\bfseries 26} (2006) 155--173}.

\bibitem{2015NatCo...6.6783B}
M.~{Bustamante}, P.~{Baerwald}, K.~{Murase} and W.~{Winter}, \emph{{Neutrino
  and cosmic-ray emission from multiple internal shocks in gamma-ray bursts}},
  \href{https://doi.org/10.1038/ncomms7783}{\emph{Nature Communications}
  {\bfseries 6} (Apr., 2015) 6783},
  [\href{https://arxiv.org/abs/1409.2874}{{\ttfamily 1409.2874}}].

\bibitem{2022ApJ...939..116A}
R.~{Abbasi}, M.~{Ackermann}, J.~{Adams}, J.~A. {Aguilar}, M.~{Ahlers},
  M.~{Ahrens} et~al., \emph{{Searches for Neutrinos from Gamma-Ray Bursts Using
  the IceCube Neutrino Observatory}},
  \href{https://doi.org/10.3847/1538-4357/ac9785}{\emph{\apj} {\bfseries 939}
  (Nov., 2022) 116}, [\href{https://arxiv.org/abs/2205.11410}{{\ttfamily
  2205.11410}}].

\bibitem{doi:10.1126/science.abg3395}
null null, R.~Abbasi, M.~Ackermann, J.~Adams, J.~A. Aguilar, M.~Ahlers et~al.,
  \emph{Evidence for neutrino emission from the nearby active galaxy ngc 1068},
  \href{https://doi.org/10.1126/science.abg3395}{\emph{Science} {\bfseries 378}
  (2022) 538--543},
  [\href{https://arxiv.org/abs/https://www.science.org/doi/pdf/10.1126/science.abg3395}{{\ttfamily
  https://www.science.org/doi/pdf/10.1126/science.abg3395}}].

\bibitem{giommi2020dissecting}
P.~Giommi, T.~Glauch, P.~Padovani, E.~Resconi, A.~Turcati and Y.~Chang,
  \emph{Dissecting the regions around icecube high-energy neutrinos: growing
  evidence for the blazar connection}, {\emph{Monthly Notices of the Royal
  Astronomical Society} {\bfseries 497} (2020) 865--878}.

\end{thebibliography}\endgroup

\end{document}